\documentclass{article}
\usepackage{spconf,amsmath,graphicx}
\usepackage[normalem]{ulem}
\usepackage{graphicx,psfrag,amsmath,amsthm,amssymb}
\usepackage{url,color}
\usepackage{algorithm,algorithmic}
\graphicspath{{grf/}}
\usepackage[square,sort,comma,numbers]{natbib}
\usepackage{sidecap}

\newcommand{\X}{\ensuremath{\mathbf{X}}}

\newcommand{\x}{\ensuremath{\mathbf{x}}}
\newcommand{\y}{\ensuremath{\mathbf{y}}}
\newcommand{\z}{\ensuremath{\mathbf{z}}}

\newcommand{\bbR}{\ensuremath{\mathbb{R}}}

{%
\begin{list}{#1}{
\vspace{-\topsep}
\vspace{-\partopsep}
\setlength{\itemindent}{0cm}
\setlength{\rightmargin}{0cm}
\setlength{\listparindent}{0cm}
\settowidth{\labelwidth}{#1}
\setlength{\leftmargin}{\labelwidth}
\addtolength{\leftmargin}{\labelsep}
\setlength{\itemsep}{0cm}
}%
}%
{%
\end{list}
\vspace{-\topsep}
\vspace{-\partopsep}
}

{\begin{enumerate}%
}%
{\end{enumerate}}

\def\x{{\mathbf x}}

\title{End-to-End Auditory Object Recognition via Inception Nucleus}
\name{Mohammad Ebrahimpour$^{1,3}$, Timothy Shea$^3$,
Andreea Danielescu$^{3}$, David Noelle$^{1,2}$, Chris Kello$^2$}
\address{Electrical Engineering and Computer Science, UC Merced~$^1$,\\ Cognitive and Information Sciences, UC Merced~$^2$,\\ Accenture Labs~$^3$}
\begin{document}
%
\maketitle
\begin{abstract}
Machine learning approaches to auditory object recognition are traditionally based on engineered features such as those derived from the spectrum or cepstrum. More recently, end-to-end classification systems in image and auditory recognition systems have been developed to learn features jointly with classification and result in improved classification accuracy. In this paper, we propose a novel end-to-end deep neural network to map the raw waveform inputs to sound class labels. Our network includes an ``inception nucleus" that optimizes the size of convolutional filters on the fly that results in reducing
engineering efforts dramatically. Classification results compared favorably against current state-of-the-art approaches, besting them by 10.4 percentage points on the Urbansound8k dataset. Analyses of learned representations revealed that filters in the earlier hidden layers learned wavelet-like transforms to extract features that were informative for classification.
\end{abstract}
\section{Introduction}
\label{s:introduction}
\vspace{-0.3cm}
Deep Convolutional Neural Networks (CNNs) have proven effective in learning to classify large sets of categories when given very large numbers of training examples~\cite{vgg,resnet}. One of the advantages of deep CNNs in object recognition is their ability to learn useful features in an end-to-end manner by mapping raw data, such as RGB pixels, to class labels.\\
In contrast, auditory object recognition is typically implemented based on engineered features~\cite{survey,logmel}. One of the most powerful types of engineered representation for speech recognition tasks is based on the mel-frequency cepstrum~\cite{log-mel}, which is basically the discrete cosine transform of the windowed spectra.
Researchers have used such engineered features as inputs to 
CNNs for audio classification tasks, such as Automatic Speech Recognition (ASR)~\cite{ASR} and music
analysis~\cite{music}. In these cases, CNNs are typically applied
to two-dimensional feature maps created by arranging the log-mel cepstral features of each frame along the time axis. This feature map creates locality in both time and frequency domains~\cite{10}, which means that the machine learning problem can be framed as an image classification problem.\\
However, cepstral features were designed specifically for speech recognition and may not be optimal for other types of audio classification tasks. More generally, pre-engineered features will be tailored to whatever the problem is to be solved, which means they may not be readily 
transferred to other problem domains. Another potential problem with engineered features is that they must be computed as inputs to the classification system, such as a deep learning convolutional network. On-line computation of spectral or cepstral features can be costly in terms of time and power, especially for edge computing applications that do not have access to cloud computing servers.
\begin{table*}[t]
\label{t:net}
\caption{Our proposed deep neural networks architectures. Each column belongs to a network. The third row indicated number of parameters. The convolutional layer parameters are denoted as ``conv (1D or 2D),(number of channels),(kernel size),(stride)." Layers with batch normalization are denoted with BN.}%
    \centering
    \scalebox{0.78}{
\begin{tabular}{|c|c|c|c|}
     \hline
     \multicolumn{4}{|c|}{Inception Nucleus Nets Configurations} \\
     \hline
     Inception & Inception-FA & Inception-FI & Inception-BN \\
     \hline
     289 K & 789 K& 479 K & 292 K \\
     \hline
     \multicolumn{4}{|c|}{Input ($32000 \times 1$)} \\
     \hline
     \multicolumn{2}{|c|}{Conv1D,32,80,4}&Inception Nucleus: &Conv1D,32,80,4 with BN \\
     \multicolumn{2}{|c|}{}&Conv1D,32,60,4& \\
     \multicolumn{2}{|c|}{}&Conv1D,[32,80,4]$\times 2$&\\
     \multicolumn{2}{|c|}{}&Conv1D,[32,100,4]$\times 2$& \\
     \hline
     Inception Nucleus:&Inception Nucleus: &Inception Nucleus:&Inception Nucleus: \\
     Conv1D,64,4,4&Conv1D,64,20,4&Conv1D,64,4,4&Conv1D,64,4,4 - BN\\
     Conv1D,[64,8,4]$\times 2$&Conv1D,[64,40,4]$\times 2$&Conv1D,[64,8,4]$\times 2$&Conv1D,[64,8,4]$\times 2$-BN\\
     Conv1D,[64,16,4]$\times 2$&Conv1D,[64,60,4]$\times 2$&Conv1D,[64,16,4]$\times 2$&Conv1D,[64,16,4]$\times 2$-BN \\
     \hline
     \multicolumn{4}{|c|}{Max Pooling 1D, 64,10,1} \\
     \hline
     \multicolumn{4}{|c|}{Reshape (put the channels first)}\\
     \hline
     \multicolumn{3}{|c|}{Conv2D,32,$3\times 3$,1}&Conv2D,32,$3\times 3$,-BN\\
     \hline
     \multicolumn{4}{|c|}{Max Pooling 2D,32,$2\times 2$,2}\\
     \hline
     \multicolumn{3}{|c|}{Conv2D,64,$3\times 3$,1}&Conv2D,64,$3\times 3$,1-BN\\
     \multicolumn{3}{|c|}{Conv2D,64,$3\times 3$,1}&Conv2D,64,$3\times 3$,1-BN\\
     \hline
     \multicolumn{4}{|c|}{Max Pooling 2D,64,$2\times 2$,2}\\
     \hline
     \multicolumn{3}{|c|}{Conv2D,128,$3\times 3$,1}&Conv2D,128,$3\times 3$,1-BN\\
     \hline
     \multicolumn{4}{|c|}{Max Pooling 2D,128,$2\times 2$,2}\\
     \hline
     \multicolumn{3}{|c|}{Conv2D,10,$1\times 1$,1}&Conv2D,10,$1\times 1$,1-BN\\
     \hline
     \multicolumn{4}{|c|}{Global Average Pooling} \\
     \hline
     \multicolumn{4}{|c|}{Softmax} \\
     \hline
\end{tabular}}
\end{table*}
More recently, researchers have developed deep learning networks that take raw waveforms as input, rather than using pre-engineered features. This approach is known as end-to-end audio classification. For instance, Dai et~al.\ proposed five CNNs with different architectures and a varying number of parameters~\cite{day_very}. They achieved impressive accuracy on the Urbansound8k dataset~\cite{day_very}. Tokozume and Harada proposed EnvNet which is an 8-layer neural network that takes the raw waveform as input, but requires careful selection of hyperparameters to choose appropriately sized kernels~\cite{envnet}. AclNet~\cite{aclnet} is another end-to-end CNN architecture, inspired by MobileNet~\cite{mobilenet} because of its computational efficiency. AclNet achieved human-level accuracy for the ESC50 dataset with only 155k parameters and 49.3 million multiply-adds per second~\cite{aclnet}. Finally, Ravanelli and Benjio proposed speaker recognition network based on raw wavforms~\cite{sincNet}.

\vspace{0.15cm}\noindent 
\textbf{Relation to prior work.}
We present a deep CNN that learns to classify broad categories of sounds directly from raw audio waveforms. In comparison to previous end-to-end audio classification efforts~\cite{day_very,envnet,aclnet}, we make use of a novel combination of 1D and 2D convolutional layers, and, most importantly, ``inception nucleus'' layers. The inception nucleus approach, described in Section~\ref{s:method}, reduces sensitivity to prespecified filter sizes by depending on adaptation during learning. In comparison to prior work, the proposed method also greatly reduces the number of parameters while outperforming current state-of-the-art networks on the urbansound8k dataset by 10.4 percentage points. Thus, our CNN is a strong candidate for low-power always-on sound classification applications. In addition, we analyze the learned representations, using visualizations to reveal wavelet-like transforms in early layers, supporting deeper representations that are discriminative and meaningful, even with reduced dimensionality.
\section{Proposed Method}
\label{s:method}
\vspace{-0.3cm}
\begin{figure}[t]
  \centering
  \begin{tabular}{c@{\hspace{0ex}}}
    \includegraphics[width=0.8\linewidth]{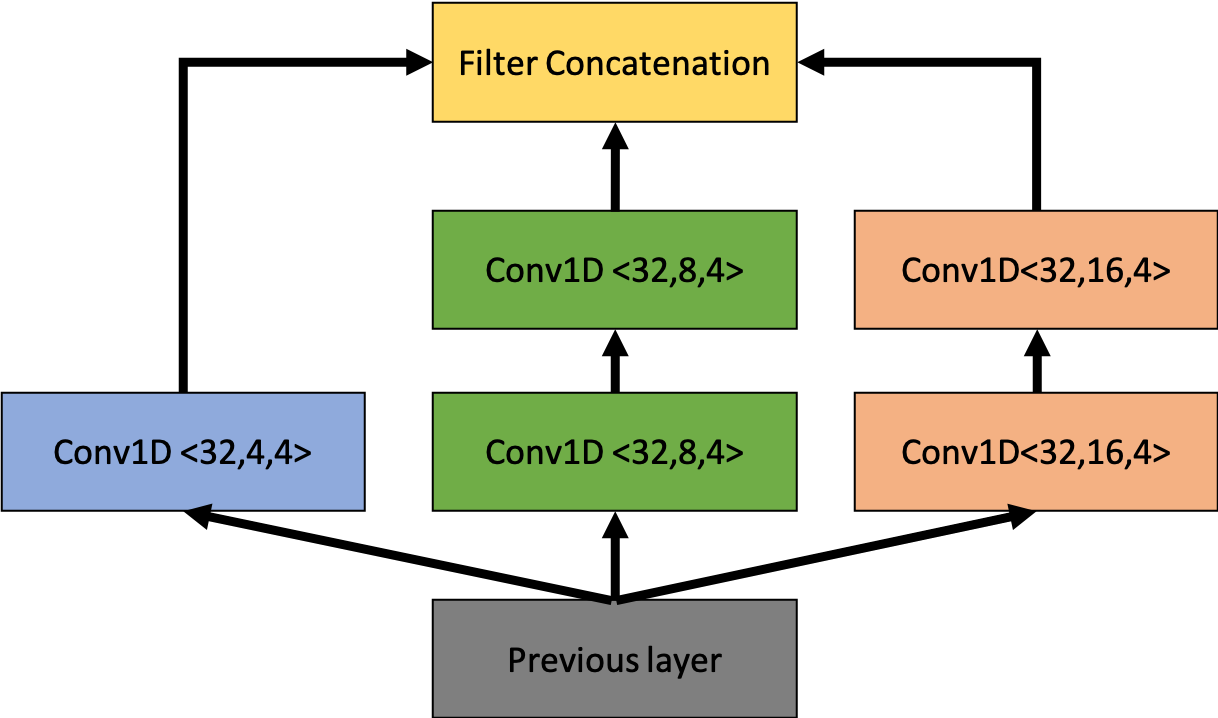}
  \end{tabular}
  \caption{Inception nucleus. The input comes from the previous layer and is passed to the 1D convolutional layers with kernel sizes of 4, 8, and 16 to capture a variety of features. The convolutional layer parameters are denoted as ``conv1D,(number of channels),(kernel size),(stride).'' All of the receptive fields are concatenated channel-wise in the concatenation layer.}
  \label{f:inception}
\end{figure} 
Our proposed end-to-end neural network takes time-domain waveform data --- not engineered representations --- and processes it through several 1D convolutions, the inception nucleus, and 2D convolutions to map the input to the desired outputs. The details of the  proposed architectures are described in Table~\ref{t:net}. The overall design can be summarized as follows:

\vspace{0.15cm}
\noindent\textbf{Fully Convolution Network.} We propose an inception nucleus convolution layer that contains a series of 1D convolutional layers followed by nonlinearities (i.e.,
ReLU layer) to reduce the sensitivity of the architecture to kernel size. Convolutional networks are well-suited for audio signals for the following reasons. First, similar to images, we desire our network to be translation invariant to reduce the number of parameters efficiently. Second,
convolutional networks allow us to stack layers, which gives us the opportunity to detect higher-level concepts through a series of lower-level detectors. We used Global Average Pooling (GAP) in our architectures to aggregate the spatial information in the last convolutional layer and map this information onto class labels. GAP greatly reduces the number of parameters to make the network relatively light to implement.

\vspace{0.15cm}
\noindent \textbf{Variable Length Input/Output.} Since sound can vary in temporal length, we want our network to handle variable-length inputs. To do this, we use a fully convolutional network. As convolutional layers are invariant to location, we can convolve each layer based on the length of the input.\\
The input layer to our network is a 1D array, representing the audio waveform, which is denoted as $\X \in \bbR^{32000 \times 1}$, since the audio files are about 4 seconds, and the sampling rate was set to be $8\ \mbox{\it kHz}$. The network is designed to learn a set of parameters, $\omega$, to map the input to the prediction, $\hat{Y}$, based on nested mapping functions, given by Eq~\ref{e:net}. 
\begin{equation}
\label{e:net}
\hat{Y} = F(\X|\omega) = f_k(...f_2(f1(\X|\omega_1)|\omega_2)|...\omega_k)
\end{equation}
where $k$ is the number of hidden layers and $f_i$ is a typical convolution layer followed by a pooling operation. 

\vspace{0.15cm}
\noindent \textbf{Inception Nucleus Layer.}
We propose the use of an inception nucleus to produce a more robust architecture for sound classification. This approach also makes the architecture less sensitive to idiosyncratic variance in audio files. A schematic representation of the inception nucleus appears in Fig~\ref{f:inception}. The inputs to the inception nucleus are the feature maps of the previous layer. Then, three 1D convolutions with different kernels are applied to the inputs to capture a variety of features. We test the following kernel sizes in our experiments: ${4,8,16,20,40,60,80,100}$. (See Section~\ref{s:experiments}.) 
After obtaining the feature maps from our convolutional layers, we concatenate the receptive fields in a channel-wise manner. 

\vspace{0.15cm}
\noindent \textbf{Reshape.}
After applying 1D convolutions on the waveforms and obtaining low-level features, the feature map, $L$, will be $\in \bbR^{1 \times m \times n}$. We can treat $L$ as a grayscale image with width=$m$, height=$n$, and channel=$1$. For simplicity, we transpose the tensor $L$ to $L^{\prime} \in \bbR^{m \times n \times 1}$. From here, we apply normal 2D convolutions with the VGG standard kernel size of $3 \times 3$ and stride = 1~\cite{vgg}. Also, the pooling layers have kernel sizes = $2 \times 2$ and stride = 2. We also implemented the inception nucleus with batch normalization to analyze the effect of batch normalization on our approach, as explained in Section~\ref{s:experiments}.

\vspace{0.15cm}
\noindent \textbf{Global Average Pooling (GAP).}
In the last convolutional layer we compute GAP to aggregate the most abstract features over the spatial dimensions and reduce the number of outputs to class labels. We use GAP instead of max pooling to reduce the number of parameters and avoid adding fully connected layers at the end of the network. It has been noted in the computer vision literature that aggregating features across spatial locations and channels keeps important information while reducing the number of parameters~\cite{ebrahimpour,CAM}. We intentionally did not use fully connected layers with a softmax activation function to avoid overfitting, since fully connected layers greatly increase the number of parameters. GAP was implemented as follows:
\begin{equation}
GAP_c = \frac{1}{w\times h} \sum_{i,j}A(i,j,c)
\end{equation}
where $w,h,c$ are width, height, and channel of the last feature map ($A$). 
\section{Experimental Results}
\label{s:experiments}
\vspace{-0.3cm}
We tested our network on the UrbanSound8k dataset which contains $10$ kinds of environmental sounds in urban areas, such as drilling, car horn, and
children playing~\cite{urbands}. The dataset consists of $8,732$ audio clips of $4$ seconds or less, totalling $9.7$ hours. We padded zeros to the samples that were less than 4 seconds.
To speed computation, the audio waveforms were down-sampled to $8\ \mbox{\it kHz}$ and standardized to zero mean and unit variance. We shuffled the training data to enhance variability in the training set.\\
We trained the CNN models using the Adam~\cite{adam} optimizer, a variant of stochastic gradient descent that adaptively tunes the step size for each dimension. We used glorot weight initialization~\cite{glorot} and trained each model with batch size $32$ for up to $300$ epochs until convergence.\\
To avoid overfitting, all weight parameters were penalized by their $\ell_2$ norm, using a $\lambda$ coefficient of $0.0001$. Our models were implemented in Keras~\cite{keras} and trained using a GeForce GTX 1080 Ti GPU.\\ %
\begin{table}[t]
\caption{Accuracy of different approaches on the UrbanSound8k dataset. The first column indicates the name of the method, the second column is the accuracy of the model on the test set, the third column reveals the number of parameters. It is clear that our proposed method has the fewest number of parameters and achieves the highest test accuracy.}
\centering
 \begin{tabular}{c c c c||} 
 \hline
Model& Test& $\#$ Parameters\\ [0.5ex] 
 \hline\hline
M3-fc~\cite{day_very} &46.82$\%$& 129M\\
M5-fc~\cite{day_very} &62.76$\%$ &18M\\
M11-fc~\cite{day_very} &68.29$\%$& 1.8M\\
M18-fc~\cite{day_very} &64.93$\%$& 8.7M\\
M3-Big~\cite{day_very}& 57.55$\%$&0.5M \\
RCNN~\cite{sang_rec2018}& 71.68$\%$& 3.7M\\
ACLNet~\cite{aclnet} &65.32$\%$& 2M\\
EnvNet-v2~\cite{tokozume2017} &78$\%$& 101M\\
PiczakCNN~\cite{piczak2015}& 73$\%$& 26M \\
VGG~\cite{pons2019}&70$\%$& 77M\\
\hline
\textbf{Inception Nucleus-BN (Ours)} & \textbf{83.2$\%$} & \textbf{292K}\\
\textbf{Inception Nucleus-FA (Ours)} & \textbf{70.9$\%$} & \textbf{789K}\\
\textbf{Inception Nucleus-FI (Ours)} & \textbf{75.3$\%$} & \textbf{479K}\\
\textbf{Inception Nucleus (Ours)} & \textbf{88.4$\%$} & \textbf{289K}\\[0.5ex] 
 \hline
\end{tabular}
\label{t:perf}
\end{table}
Table~\ref{t:perf} provides classification performance on the testing set along with numbers of parameters used for the Urbansound8k dataset. The table shows that our CNN outperformed other methods in terms of test classification accuracy, with the fewest number of parameters. Preliminary simulations revealed that fully connected layers at the end of the network caused overfitting due to an explosion in the number of weight parameters. These preliminary results led us to use a fully convolutional network with a reduced number of parameters.\\
\begin{figure}[h]
  \centering
  \begin{tabular}{c@{\hspace{0ex}}c@{\hspace{0ex}}c@{\hspace{0ex}}}
    \includegraphics[width=0.3\linewidth]{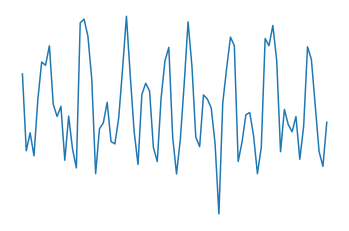}&
    \includegraphics[width=0.3\linewidth]{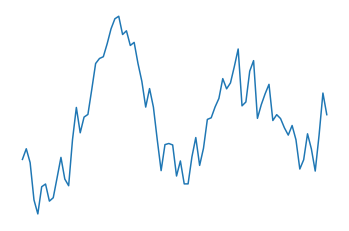}&
    \includegraphics[width=0.3\linewidth]{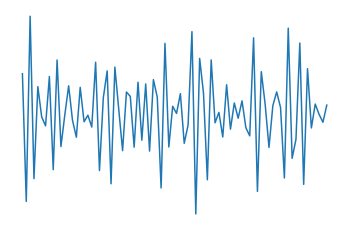}\\
    \includegraphics[width=0.3\linewidth]{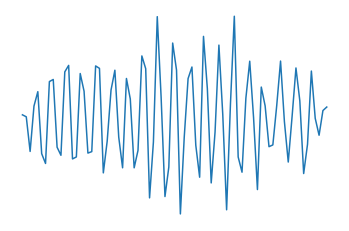}&
    \includegraphics[width=0.3\linewidth]{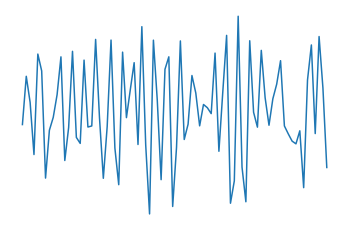}&
    \includegraphics[width=0.3\linewidth]{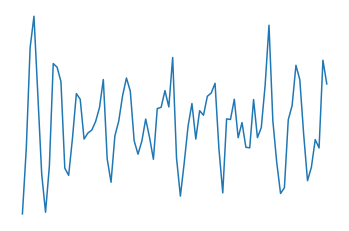}\\
  \end{tabular}
  \caption{Illustrating 3 filters in the first convolutional layer. The visualization indicates that learned representations in the early layers implemented wavelet-like audio filters.}
  \label{f:filters}
\end{figure} 
\begin{figure}[h]
  \centering
  \begin{tabular}{c@{\hspace{0ex}}}
        \includegraphics[width=0.9\linewidth]{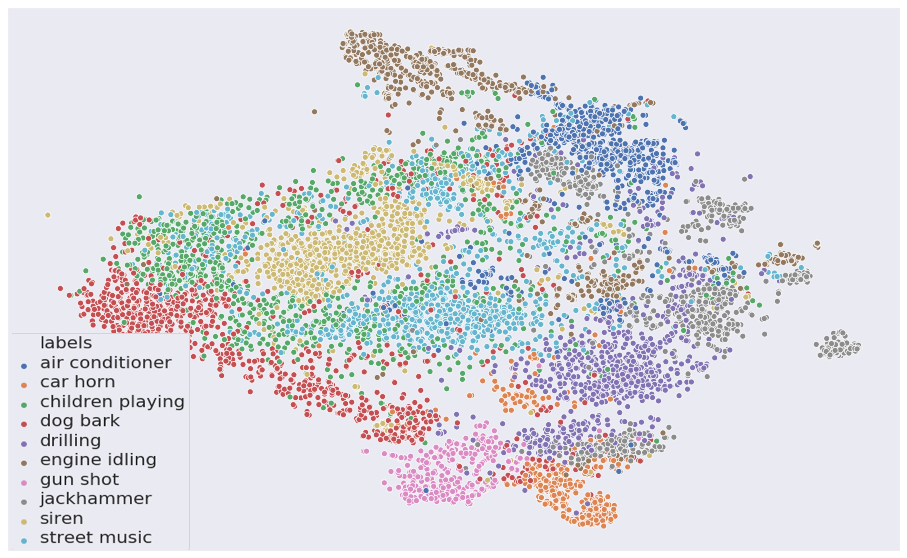}
  \end{tabular}
  \caption{Illustration of the top two components of the t-SNE of the last convolutional layer.}
  \label{f:tsne}
\end{figure} 
We note that the deeper networks (M5, M11, and M18) can improve performance if their architectures are well-designed. However, our inception nucleus model is $88.4\%$ accurate, which outperforms the reported test accuracy of CNNs on spectrogram input using the same dataset by a large margin~\cite{piczak2015}. Also, inception nucleus-FI achieves very good results in terms of both accuracy and number of parameters. This result suggests that if we let the network learn useful features for the desired task in the convolutional layers, recognition performance and generalization is improved over pre-engineered features.

\vspace{0.15cm}
\noindent\textbf{Kernel Analysis.}
We also analyzed the learned kernels of our Inception Nucleus model in the very first layer of our neural network. Interestingly, the network learns wavelet transforms at the first convolutional layer, as has been found by other researchers~\cite{soundnet,ACISCA}. Some of those filters are illustrated in Fig~\ref{f:filters}.

\vspace{0.15cm}
\noindent\textbf{Representation Analysis.}
To better understand the learned representations in the Inception Nucleus model, we extracted features from the last convolutional layer (before applying GAP) and applied t-SNE to reduce the dimensionality to two~\cite{tsne}. The results, shown in Fig.~\ref{f:tsne}, suggest that the network learned meaningful and discriminative features, as the different classes are fairly well distinguished from each other.

\vspace{0.15cm}
\noindent\textbf{Depth Analysis.}
We found that deeper networks with larger numbers of parameters were less generalizable as indicated by poorer performance on the test set. For example, M18 has $8.7M$ parameters (see Table~\ref{t:perf}) but only achieves $64.93\%$ accuracy, compared with our inception nucleus network which achieves $88.4\%$ by only having $289 K$ parameters. This finding runs counter to results from the image recognition literature, in which deeper networks tend to perform better than shallower ones~\cite{resnet,densenet,senet}. The observed detriment of additional hidden layers may be attributable to the limited number of training examples, which can be tested in future studies with larger datasets.

\vspace{0.15cm}
\noindent\textbf{Kernel Size Analysis.}
Dai et~al.~\cite{day_very} found that smaller kernel sizes are insufficient to capture the necessary bandpass filter characteristics in the earlier convolutional layers. Our results indicate that, with the Inception Nucleus-FS, large kernel sizes (e.g. $60,80,100$) are more effective in the first convolutional layer. By contrast, large kernel sizes in the second layer reduce performance substantially (e.g., using the Inception Nuclus-FA with large kernels in the second layer decreased performance by 13 points). We conclude that a larger inception nucleus is more suitable for the first layer, with smaller kernels in later convolutional layers. 

\vspace{0.15cm}
\noindent\textbf{Batch Normalization.}
We tested whether batch normalization (BN) improves performance in our CNN, as it can for very deep neural networks. Without BN, our inception nucleus achieves $88.4\%$ accuracy while with BN it achieves $83.2\%$. The slight decrease in accuracy using BN may have been observed because our CNNs did not have enough layers to show the advantage of BN. 
\section{Conclusion}
\vspace{-0.3cm}
In this study, we developed, optimized, and tested CNNs up to 13 layers deep that used an inception nucleus to overcome problems with choosing kernel sizes. The CNNs were trained to perform end-to-end sound classification, and they were benchmarked against the Urbansound8K dataset. Results from our networks, compared with competitors, showed better performance with fewer parameters --- up to $88.4\%$ accuracy using only $289 K$ parameters. The ability to perform end-to-end computations effectively using so few parameters may be useful for edge computing applications, especially with optimized hardware, such as neuromorphic implementations of deep networks~\cite{neuromorphic}.
Our results indicate that end-to-end computation does not detract from performance by forgoing cepstral or spectral features. To the contrary, our networks outperformed competitors that used log-mel spectrogram inputs~\cite{piczak2015}. Visualizations of kernels learned in the earliest layer revealed wavelet-like transforms that build up to more abstract and discriminating learned features in deeper layers. In summary, we have demonstrated effective end-to-end sound classification with an efficient deep learning network. 
\section*{Acknowledgements}
\vspace{-0.3cm}
This research was supported in part by a gift from Accenture Labs (LLP) to Cognitive and Information Sciences at the University of California, Merced (PI Kello). 

\bibliographystyle{IEEEbib}
\bibliography{strings,refs}
\end{document}